%% file: zjets_plb_resubmission.tex
\documentclass[aps,prl,showpacs,twocolumn,groupedaddress]{revtex4}
\usepackage{graphicx}
\usepackage{dcolumn} 
\usepackage{bm}      
\usepackage{amssymb} 
\usepackage{color}

\def \pt       {\ensuremath{ p_T}}
\def \ptz      {\ensuremath{ p_{T,Z}}}
\def \ptzTwo   {\ensuremath{ p_{T,Z}^2}}
\def \ptjet    {\ensuremath{ p_{T}^{\mathrm{jet}} }}

\def \ifb    {\ensuremath{ \rm fb^{-1}                          }}

\newcommand{\gev}{\mbox{\,\rm Ge\kern -0.1emV}}
\newcommand{\mev}{\mbox{\,\rm Me\kern -0.1emV}}
\newcommand{\Eslash}{\mbox{$\rm E \kern-0.6em\slash$                }}

\newcommand{\dzero}{D0}
\def \ppbar    {\ensuremath{ p\bar{p}}}
\def \qqbar    {\ensuremath{ q\bar{q}}}
\def \pt    {\ensuremath{p_T}}
\def \vjets {$V+$\hspace{.5mm}jets}

\begin{document} 

\hspace{5.2in}
\mbox{FERMILAB-PUB-09-066-E}

\title{Measurements of differential cross sections of {\boldmath$
Z/\gamma^*+$}jets{\boldmath$+X$} events in {\boldmath$p\bar{p}$}
 collisions at {\boldmath$\sqrt{s}=1.96$}~TeV}

\input list_of_authors_r2.tex  
\date{March 10, 2009}

\begin{abstract}

We present cross section measurements for $Z/\gamma^*+$jets$+X$
production, differential in the transverse momenta of the three
leading jets. The data sample was collected with the
\dzero~detector at the Fermilab Tevatron \ppbar~collider at a 
center-of-mass energy of $1.96$ TeV and corresponds to an integrated
luminosity of $1$~\ifb. Leading and next-to-leading order perturbative
QCD predictions are compared with the measurements, and agreement is
found within the theoretical and experimental uncertainties. We also
make comparisons with the predictions of four event generators. 
Two parton-shower-based generators show significant shape and
normalization differences with respect to the data.
In contrast, two generators combining tree-level matrix elements with
a parton shower give a reasonable description of the the shapes
observed in data, but the predicted normalizations show significant
differences with respect to the data, reflecting large scale
uncertainties. For specific choices of scales, the normalizations for
either generator can be made to agree with the measurements.

\end{abstract} 

\pacs{12.38.Qk, 13.85.Qk, 13.87.-a}

\maketitle

The production of jets in association with vector bosons (\vjets) in
hadron collisions is an important process in quantum chromodynamics
(QCD) and is a significant source of background for many standard
model measurements (e.g., $t\bar{t}$ production) and in searches for
new phenomena (e.g., supersymmetry). Such measurements at the Fermilab
Tevatron collider and the CERN Large Hadron Collider (LHC) rely on
accurate descriptions of \vjets~production by particle-level event
generators. These models require validation with measurements of the
properties of the \vjets~system, especially as a function of jet
multiplicity.

In this Letter, we present new precision measurements of differential
cross sections for the production of $Z/\gamma^*+$jets$+X$ in
$p\bar{p}$ collisions at a center-of-mass energy of $1.96$~TeV. Cross
sections are presented in bins of the transverse momentum (\pt) of the
$N^{\rm th}$ jet in events containing at least $N=1$, $2$,~or $3$~jets
and are normalized to the measured inclusive $Z/\gamma^*$ cross
section to reduce uncertainties. The $N$~jets are ordered in terms of
decreasing $p_T$, and the $Z/\gamma^*$ is selected via its decay into
an electron-positron pair. The data set corresponds to an integrated
luminosity of $1.04\pm 0.06~\mathrm{fb}^{-1}$~\cite{lumi}.

Previous $Z/\gamma^*+$jets measurements have focused primarily on
measuring the jet multiplicities for up to three~\cite{cdfZjets} and
four~\cite{d0ZjetMult} jets. In addition, differential distributions
have been presented for the highest-\pt~(leading) jet~\cite{d0Zjets}
and for the two leading jets~\cite{cdfZjets}. In this Letter, we
extend the measurements of differential distributions by including the
third leading jet, and by including a larger \ptjet~range. A major
focus of the Letter is a comparison of the differential
\pt~distributions to leading order (LO) and next-to-leading (NLO)
perturbative QCD (pQCD) predictions from {\sc mcfm}~\cite{mcfm}, as
well as to results from several commonly-used event generators. In
particular, we have investigated to what extent the differential
\pt~distributions for the higher jet multiplicities can be described
by parton-shower-based event generators like {\sc pythia}
\cite{pythia64} and {\sc herwig}~\cite{herwig}, and how they compare
to event generator predictions where matrix element and parton shower
merging procedures are adopted, as in {\sc
  alpgen+pythia}~\cite{alpgen} and {\sc sherpa}~\cite{sherpa}. For
each prediction, the renormalization and factorization scale
uncertainties are evaluated, and the choices of scales are modified to
achieve an improved description of the measurements. The presented
studies are of vital importance to understand the predictive power of
the various event generator models for \vjets~processes at both the
Tevatron and the LHC.

The data set was recorded using the \dzero~Run~II detector, which is
described in detail elsewhere~\cite{run2det}. Here we give a brief
overview of the most relevant components for this analysis. The
trajectory of charged particles are reconstructed using a silicon
vertex tracker and a scintillating fiber tracker located inside a
superconducting solenoidal coil that provides a magnetic field of
approximately $2$ T. The tracking volume is surrounded by a
liquid-argon and uranium calorimeter, divided into electromagnetic and
hadronic sections with a granularity of $\Delta \eta \times \Delta
\phi = 0.1 \times 0.1$, where $\eta$ is the pseudorapidity 
\cite{pseudorapidity} and $\phi$ is the azimuthal angle. The third 
layer of the electromagnetic calorimeter has a finer granularity of
$\Delta \eta \times \Delta \phi = 0.05 \times 0.05$. The calorimeter
consists of three sections, each housed in a separate cryostat, with a
central section covering $|\eta| \le 1.1$ and two end calorimeters
extending the coverage to $|\eta| \approx 4.2$. Scintillators between
the cryostats sample shower energy for $1.1 < |\eta|< 1.4$.

Electrons are identified based on their characteristic energy
deposition signature in the calorimeter, including the transverse and
longitudinal shower profiles. In addition, a reconstructed track must
point to the energy deposit in the calorimeter, and the momentum of
the track and the calorimeter energy must be consistent. Rejection
against background from photons and jets is achieved with a likelihood
discriminant which uses calorimeter and tracking information
\cite{emId}.

Calorimeter jets are reconstructed using the \dzero~Run~II iterative
seed-based cone jet algorithm~\cite{midPoint}, using a
splitting/merging fraction of 0.5 and a cone radius ${\cal
R}\nolinebreak =\nolinebreak \sqrt{ ( \Delta y )^2 + (\Delta \phi )^2
} = 0.5$, with $y$ being the rapidity~\cite{rapidity}. The input
objects are clusters of energy deposited in the calorimeter. Rejection
of jets arising from electronic noise in the calorimeter is achieved
by using quality and jet shape cuts. The reconstructed jets are
corrected for the calorimeter response, instrumental out-of-cone
showering effects, and additional energy from multiple
\ppbar~interactions and previous beam crossings. These corrections
were derived by exploiting the \pt~balance in $\gamma+$jet and dijet
events. The measurements are performed for jets with $\pt > 20$~GeV
and $|\eta| < 2.5$.

Events were required to pass single or dielectron trigger
requirements, and to contain two electron candidates with opposite
sign electric charge, $\pt > 25$~GeV, $|\eta|<1.1$ or
$1.5<|\eta|<2.5$, and a dielectron mass ($M_{ee}$) satisfying $65 <
M_{ee} < 115$~GeV. A total of 65,759 events pass the selection before
background subtraction. Of these, 8,452/1,233/167 events have 1/2/3
jets or more, with \ptjet~above 20~GeV. The efficiency for the trigger
requirements to be satisfied by $Z/\gamma^*(\rightarrow e^+e^-)$
events which fulfill the other selection criteria was found to be
$\sim 100\%$, independent of the number of jets in the event.

Backgrounds arising from events which contain two real electrons
(e.g.,~$WW$, $t\bar{t}$, and $Z/\gamma^*$($\rightarrow\tau^+\tau^-$))
were estimated using event samples generated with {\sc pythia} v6.323,
with the underlying event model configured using Tune
A~\cite{tuneA}. The events were passed through a {\sc
  geant}-based~\cite{geant} simulation of the detector response, and
each event sample was normalized to higher-order theoretical
predictions~\cite{mcfm,ttbarCrossSection} before being subtracted from
the measured data. For the inclusive data sample, the estimated sum of
backgrounds arising from events containing two real electrons is $\sim
0.2\%$. Of the background sources, only $t\bar{t}$ production has an
average jet multiplicity which is significantly larger than that of
the signal process. As a result, $t\bar{t}$ is the dominant background
for events containing two or more jets, contributing up to $6\%$
($3\%$) to the measured data for large values of \pt\ of the second
(third) jet. The sum of all backgrounds containing two real electrons
is estimated to be less than one third of the statistical uncertainty
of the measured data in all bins of the measurement. By studying a
sample of same-charge dielectron events, backgrounds arising from
events with one or more mis-reconstructed electrons (e.g., $W+$jets or
multijet events) were found to be below $1\%$ in the inclusive
$Z/\gamma^*(\rightarrow e^+e^-)$ sample. For the signal samples
containing at least $1, 2$, or $3$ jets, no statistically significant
contribution from $W+$jets or multijet events was observed. For
$W+$jets, this was confirmed independently by using an event sample
generated with {\sc pythia}, yielding a contribution to the signal
samples at the $0.1\%$ level.

The corrections of the reconstructed \ptjet~spectra to the particle
level~\cite{particleLevel} were determined using an event sample
generated with {\sc alpgen} v2.05 + {\sc pythia} v6.325 using Tune
A. The events were passed through a {\sc geant}-based simulation of
the detector response. The simulated events were overlaid with data
events from random bunch crossings to reproduce the effects of
detector noise and additional \ppbar~interactions. Particle-level jets
were defined through the \dzero~Run~II iterative seed-based cone jet
algorithm with ${\cal R}=0.5$ using a splitting/merging fraction of
0.5 and were required to satisfy $|y|<2.5$. The input objects were all
stable particles except the two electrons defining the dielectron
system and any photons in a cone of ${\cal R}=0.2$ around the two
electrons. The latter requirement excludes most photons associated
with the $Z/\gamma^*$ decays, described in the simulation as QED
final-state radiation.

In the simulated sample, the ratio of the reconstructed
\ptjet~spectrum to the particle level spectrum defines the product of
the efficiency of the detector ($\epsilon_{\rm sim}$) and its
geometrical acceptance ($A_{\rm sim}$). The product $(\epsilon \times
A)$ also includes the impact of the detector resolution on the
reconstructed spectrum. To achieve agreement between $(\epsilon
\times A)_{\rm sim}$ and the corresponding quantity in data,
$(\epsilon \times A)_{\rm data}$, the simulated event sample was
modified in two steps. First, the simulated event sample was corrected
so that its object identification efficiencies, energy scales, and
energy resolutions correspond to those measured in data. Next, the
shapes of the \ptjet~spectra in the simulated sample were reweighted
at the particle level so that agreement with data was obtained at the
reconstructed level. These two steps are explained in more detail
below.

The single electron identification efficiency was measured in data for
inclusive $Z/\gamma^*(\rightarrow e^+e^-)+X$ events
\cite{emEff}, giving an efficiency, including statistical uncertainty,
of $\epsilon_{\rm data}^{e} = 0.77\pm 0.01$. The electron
identification efficiency in the simulated sample, $\epsilon_{\rm
sim}^{e}$, was found to be $\approx 10\%$ higher than in data. To
adjust the simulation to correspond to data, each reconstructed
electron candidate in the simulated sample was rejected with a
probability given by $\epsilon_{\rm data}^{e} / \;\epsilon_{\rm
sim}^{e}$. The electron efficiencies were binned in $\eta_{e}$ and
$\phi_{e}$. As a cross check, the analysis was also performed using
electron efficiencies binned in $\eta_{e}$ and $p_{T,e}$, resulting in
changes of less than $1\%$ in the final measurements. The same
correction procedure was applied for the identification efficiency for
jets, which was found to be $\epsilon_{\rm data}^{\rm jet} = 0.98\pm
0.02$ in data.

The electron identification efficiency depends on the number of jets
in an event and on their kinematics. Electron candidates with a nearby
jet are less likely to pass the electron identification criteria. To
determine if the correlation between the electron efficiencies and the
jet activity is correctly described by the simulation, two sets of
comparisons were performed. First, the minimal $\Delta{\cal R}$
separation between each electron and all jets in $N$-jet events
($N=0,1,2,3$) in the simulated sample were compared with data. Second,
the amount of soft QCD radiation not clustered into jets was studied
using tracks which were not associated with any reconstructed electron
or jet. The multiplicity and \pt~sum of all such tracks in $N$-jet
events ($N=1,2,3$), relative to those in $0$-jet events, were compared
between simulation and data. For both sets of comparisons, reasonable
agreement was found. The total uncertainty due to differences between
the simulation and the data in the dependence of the electron
identification efficiency on jet activity was estimated to be below
2\% for all cross section measurements.

Jet energy scale (JES) corrections were derived using $\gamma$+jet and
dijet events, assuring that the reconstructed \ptjet~in $\gamma$+jet
events on average is equal to the particle-level \ptjet. The
difference in calorimeter response between quark- and gluon-initiated
jets introduces a dependence of the JES on the physics process. As a
result, the JES corrections derived using $\gamma$+jet events do not
guarantee a correct JES when applied to $Z/\gamma^*+$jet
events. However, the correction of data to the particle level depends
only on data and simulation having a common JES. The factor $(\epsilon
\times A)$ accounts for any differences between this common JES and
the correct JES since particle-level jets in the simulated sample by
definition have the proper energy scale. The \pt~balance in
$Z/\gamma^*+$jet events was used to adjust the JES in simulation to be
equal to the JES in data. The correction is $\sim ~5\%$ below $40$~GeV
and becomes negligible above $80$~GeV.

The jet energy resolution (JER) in data and simulated events was
determined using $Z/\gamma^*+$jet events, and the reconstructed jet
energies in the simulated sample were corrected to account for any
differences. The JER distorts the steeply falling jet
\pt~spectra, resulting in a net migration towards higher values of
\ptjet. This leads to the reconstructed
\ptjet~spectra being (5--15)\% higher than they would have been for a
detector with perfect jet energy resolution. The factor
$(\epsilon\times A)$ accounts for this effect, and the measurements
are fully corrected for the JER.

After adjusting the simulated detector performance to the real
detector performance, we tuned the \ptjet~spectra in the simulation to
those observed in data. For the measurement of each observable, the
simulated event sample was reweighted as a function of that observable
at the particle level to ensure agreement of the distribution of the
observable at the reconstructed level. After the reweighting,
$(\epsilon \times A)_{\rm sim}$ is equal to $(\epsilon\times A)_{\rm
data}$ within the uncertainties of the corrections applied to the
simulated sample. The particle level spectra in data equal the
reconstructed spectra in data divided by $(\epsilon \times A)_{\rm
sim}$. The total efficiency for reconstructing a
$Z/\gamma^*(\rightarrow e^+e^-)+N$-jets$+X$ event varies with
\ptjet~and is in the range 0.5--0.7 for $N=1$ and 0.4--0.5 for
$N=2,3$.

For events containing more than one jet, the $N^{\rm th}$ jet at the
particle level might not be equal to the $N^{\rm th}$ jet at the
reconstructed level. This can occur for events containing two jets
with similar values of \ptjet. The impact of this effect is part of
$(\epsilon \times A)$ as defined above, and it is correctly taken into
account if the JER is identical for simulation and data, and if two
jets have similar \ptjet~values equally often in simulation and in
data. The former is assured by the JER corrections applied to the
simulated event sample, as described above. To test if the latter is
satisfied, the ratios of the reconstructed \pt-spectrum of the $N^{\rm
th}$ jet to the spectra of the $(N-1)^{\rm th}$ jet and the
$(N+1)^{\rm th}$ jet in the simulated event sample was compared with
those observed in data. Agreement was observed within statistical
uncertainties.

The final measurements are presented with two different particle-level
selections in the mass range $65 < M_{ee} < 115$~GeV: first without
any further selections on the electrons (selection {\it a}) and
secondly requiring $\pt^{e} > 25$~GeV and $|y^{e}|<1.1$ or
$1.5<|y^{e}|<2.5$ (selection {\it b}). For each particle-level
electron selection, the \ptjet\ spectra were normalized to the
inclusive $Z/\gamma^*(\rightarrow e^+e^-)+X$ cross section measured
with the same particle-level selection. Selection {\it b} corresponds
to the kinematic range which is measured in data. Selection {\it a}
includes an extrapolation to the full range of lepton kinematics in
order to simplify direct comparisons with other measurements. The
extrapolation factor was derived from event samples generated using
{\sc sherpa} v1.1.1, {\sc alpgen} v2.13 + {\sc pythia} v6.325 using
Tune A and {\sc pythia} v6.323 using Tune A. The central value of the
extrapolation factor was taken to be the {\sc sherpa} prediction, with
the maximal deviation to the two other predictions being assigned as a
systematic uncertainty. The uncertainties due to the PDFs were
evaluated using the Hessian method~\cite{hessian} and were found to be
negligible. The extrapolation from selection {\it b} to selection {\it
  a} increases the normalized differential cross section for the
leading jet by $\approx 10\%$ below $100$~GeV and decreases it by
$\approx 25\%$ above $200$~GeV. For the second (third) jet, the
extrapolation changes the observable by less than $10\%$ ($2\%$).

\begin{table*}[!t]
\caption{The measurements of $\frac{1}{\sigma_{Z/\gamma^*}}\times
 \frac{d\sigma}{d \pt}$ for the first jet (\pt~ordered) in
  $Z/\gamma^*(\rightarrow {e^+e^-})$ events with one or more jets.}
\label{tab:results_jpt0}
\vspace{1mm}
\begin{ruledtabular}
\begin{tabular}{r@{\hspace{-0.5mm} --\hspace{-5.3mm}}lccccccc}
\multicolumn{2}{c}{}& & \multicolumn{3}{c}{$\pt^{e}>25$~GeV, $|y^{e}|<1.1$ or $1.5<|y^{e}|<2.5$,}&
\multicolumn{3}{c}{} \\
\multicolumn{2}{c}{}& & \multicolumn{3}{c}{$65<M_{ee}<115$~GeV} &
\multicolumn{3}{c}{$65<M_{ee}<115$~GeV} \\
\hline
\multicolumn{2}{c}{\pt~bin} & Bin center 
& $1 / \sigma_{Z/\gamma^*} \times d\sigma / d \ptjet $
& $\delta_{\rm stat}$ & $\delta_{\rm sys}$
& $1 / \sigma_{Z/\gamma^*} \times d\sigma / d \ptjet $
& $\delta_{\rm stat}$ & $\delta_{\rm sys}$
\\ 
\multicolumn{2}{c}{$[{\rm GeV}]$} & $[{\rm GeV}]$ 
& $[1/{\rm GeV}]$ & $(\%)$ & $(\%)$
& $[1/{\rm GeV}]$ & $(\%)$ & $(\%)$\\
\hline
\hspace{0.3mm}  20&28  & 23.7  & $6.81\times 10^{-3}$ &    1.6 &   5.9 & $7.19\times 10^{-3}$ & 1.6 &   7.0 \\
\hspace{0.3mm}  28&40  & 33.5  & $2.99\times 10^{-3}$ &    2.1 &   4.0 & $3.22\times 10^{-3}$ & 2.1 &   4.6 \\
\hspace{0.3mm}  40&54  & 46.4  & $1.23\times 10^{-3}$ &    3.1 &   3.3 & $1.37\times 10^{-3}$ & 3.1 &   3.9 \\
\hspace{0.3mm}  54&73  & 62.7  & $5.04\times 10^{-4}$ &    4.0 &   3.3 & $5.74\times 10^{-4}$ & 4.0 &   4.1 \\
\hspace{0.3mm}  73&95  & 83.1  & $2.03\times 10^{-4}$ &    5.6 &   7.6 & $2.27\times 10^{-4}$ & 5.6 &   8.2 \\
\hspace{0.3mm}  95&120 & 106.2 & $7.29\times 10^{-5}$ &    8.4 &   8.4 & $7.62\times 10^{-5}$ & 8.4 &   8.9 \\
\hspace{0.3mm} 120&154 & 135.3 & $2.64\times 10^{-5}$ &   12   &  10   & $2.54\times 10^{-5}$ & 12  &  11   \\
\hspace{0.3mm} 154&200 & 172.9 & $8.08\times 10^{-6}$ &   17   &  14   & $6.99\times 10^{-6}$ & 17  &  14   \\
\hspace{0.3mm} 200&300 & 236.9 & $7.46\times 10^{-7}$ &   42   &   25  & $5.58\times 10^{-7}$ & 42  &  25   \\
\end{tabular}
\end{ruledtabular}
\end{table*}

\begin{table*}[!t]
\caption{The measurements of $\frac{1}{\sigma_{Z/\gamma^*}}\times
 \frac{d\sigma}{d \pt}$ for the second jet (\pt~ordered) in
  $Z/\gamma^*(\rightarrow {e^+e^-})$ events with two or more
  jets.}
\label{tab:results_jpt1}
\vspace{1mm}
\begin{ruledtabular}
\begin{tabular}{r@{\hspace{-0.5mm} --\hspace{-5.3mm}}lccccccc}
\multicolumn{2}{c}{}& & \multicolumn{3}{c}{$\pt^{e}>25$~GeV, $|y^{e}|<1.1$ or $1.5<|y^{e}|<2.5$,}&
\multicolumn{3}{c}{} \\
\multicolumn{2}{c}{}& & \multicolumn{3}{c}{$65<M_{ee}<115$~GeV} &
\multicolumn{3}{c}{$65<M_{ee}<115$~GeV} \\
\hline
\multicolumn{2}{c}{\pt~bin} & Bin center 
& $1 / \sigma_{Z/\gamma^*} \times d\sigma / d \ptjet $
& $\delta_{\rm stat}$ & $\delta_{\rm sys}$
& $1 / \sigma_{Z/\gamma^*} \times d\sigma / d \ptjet $
& $\delta_{\rm stat}$ & $\delta_{\rm sys}$
\\ 
\multicolumn{2}{c}{$[{\rm GeV}]$} & $[{\rm GeV}]$ 
& $[1/{\rm GeV}]$ & $(\%)$ & $(\%)$
& $[1/{\rm GeV}]$ & $(\%)$ & $(\%)$\\
\hline
\hspace{1.5mm} 20&28 &  23.6 & $1.30\times 10^{-3}$ &   3.7 &  10   & $1.39\times 10^{-3}$ &   3.7 &  11   \\
\hspace{1.5mm} 28&40 &  33.3 & $4.23\times 10^{-4}$ &   5.7 &   5.2 & $4.51\times 10^{-4}$ &   5.7 &   6.0 \\
\hspace{1.5mm} 40&54 &  46.2 & $1.57\times 10^{-4}$ &   9.1 &   6.2 & $1.62\times 10^{-4}$ &   9.1 &   6.8 \\
\hspace{1.5mm} 54&73 &  62.3 & $4.17\times 10^{-5}$ &  15   &   8.5 & $4.20\times 10^{-5}$ &  15   &   9.0 \\
\hspace{1.5mm} 73&200& 112.9 & $2.96\times 10^{-6}$ &  22   &   7.4 & $2.82\times 10^{-6}$ &  22   &   8.0 \\
\end{tabular}
\end{ruledtabular}
\end{table*}

\begin{table*}[!t]
\caption{The measurements of $\frac{1}{\sigma_{Z/\gamma^*}}\times
 \frac{d\sigma}{d \pt}$ for the third jet (\pt~ordered) in
  $Z/\gamma^*(\rightarrow {e^+e^-})$ events with three or more
  jets.}
\label{tab:results_jpt2}
\vspace{1mm}
\begin{ruledtabular}
\begin{tabular}{r@{\hspace{-0.5mm} --\hspace{-5.3mm}}lccccccc}
\multicolumn{2}{c}{}& & \multicolumn{3}{c}{$\pt^{e}>25$~GeV, $|y^{e}|<1.1$ or $1.5<|y^{e}|<2.5$,}&
\multicolumn{3}{c}{} \\
\multicolumn{2}{c}{}& & \multicolumn{3}{c}{$65<M_{ee}<115$~GeV} &
\multicolumn{3}{c}{$65<M_{ee}<115$~GeV} \\
\hline
\multicolumn{2}{c}{\pt~bin} & Bin center 
& $1 / \sigma_{Z/\gamma^*} \times d\sigma / d \ptjet $
& $\delta_{\rm stat}$ & $\delta_{\rm sys}$
& $1 / \sigma_{Z/\gamma^*} \times d\sigma / d \ptjet $
& $\delta_{\rm stat}$ & $\delta_{\rm sys}$
\\ 
\multicolumn{2}{c}{$[{\rm GeV}]$} & $[{\rm GeV}]$ 
& $[1/{\rm GeV}]$ & $(\%)$ & $(\%)$
& $[1/{\rm GeV}]$ & $(\%)$ & $(\%)$\\
\hline

\hspace{1.5mm} 20&28 & 23.6 & $2.22\times 10^{-4}$ &   9.1 &  14   & $2.33\times 10^{-4}$ &   9.1 &  16 \\
\hspace{1.5mm} 28&44 & 34.6 & $4.40\times 10^{-5}$ &  17   &   8.4 & $4.48\times 10^{-5}$ &  17   &  11 \\
\hspace{1.5mm} 44&60 & 50.9 & $8.67\times 10^{-6}$ &  42   &  11   & $8.60\times 10^{-6}$ &  42   &  13 \\

\end{tabular}
\end{ruledtabular}
\end{table*}

The systematic uncertainties of the measurements arise from the
uncertainties of the background estimates, which were found to be
negligible, and from the uncertainties of the corrections applied to
the simulated event sample to assure that $(\epsilon \times A)_{\rm
  sim} = (\epsilon\times A)_{\rm data}$. Each correction was varied
separately within its uncertainties, and the resulting variations in
the measured \ptjet\ spectra were added in quadrature to give the
total systematic uncertainty of the measurements. The largest source
of uncertainty is the correction of the JES in the simulated event
sample to correspond to the JES of the data sample, contributing
(50--80)\% of the total systematic uncertainty of the
measurements. Additional uncertainties arise from the reweighting
function applied to the simulated event sample at the particle level
to ensure agreement with data at the reconstructed level, from the
correction of the jet energy resolution and of the jet and electron
identification efficiencies in the simulated event sample, and from
the extrapolation from selection {\it b} to selection {\it
  a}. Presenting the measurements as ratios to the inclusive
$Z/\gamma^*(\rightarrow e^+e^-)+X$ cross section cancels the
dependence on the uncertainty in the integrated luminosity of the data
set. Additionally, most of the dependence on the uncertainties in the
electron trigger and identification efficiencies also cancels.

The choice of binning for the measurements was guided by the finite
JER, which causes events to contribute to different bins at the
particle and reconstruction levels. The purity of a bin is defined as
the fraction of the simulated events which are reconstructed in the
same bin in which they were generated at the particle level. The
widths of the measurement bins were chosen so that each bin has a
purity of about $60$\%.

\begin{figure*}
\begin{center}
  \includegraphics*[width=\linewidth]{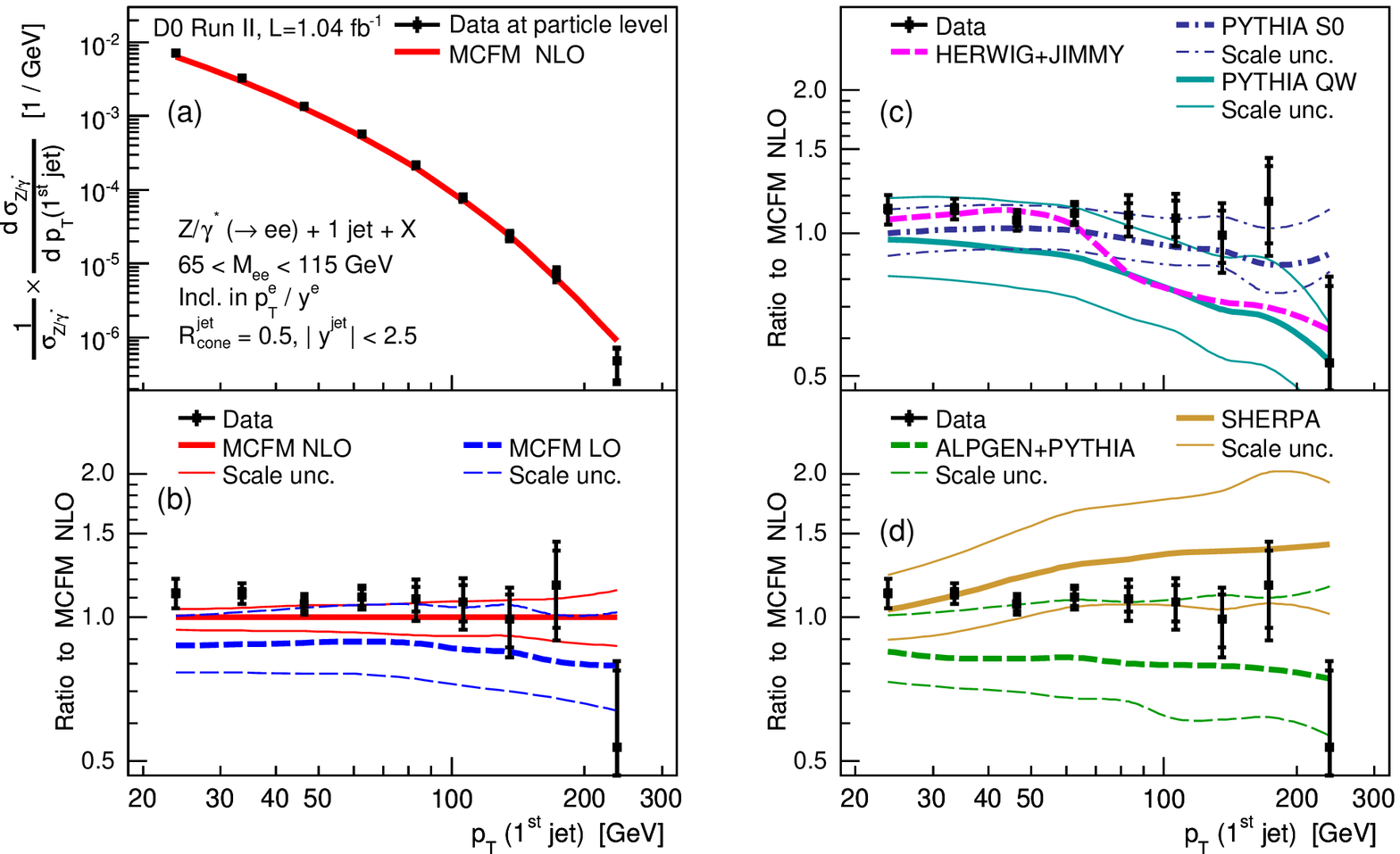}
  \caption{\label{fig1} (a) The measured distribution of
  $\frac{1}{\sigma_{Z/\gamma^*}}\times \frac{d\sigma}{d \pt({\rm
  jet})}$ for the leading jet in $Z/\gamma^*+$jet+$X$ events, compared
  to the predictions of {\sc mcfm nlo}. The ratios of data and theory
  predictions to {\sc mcfm nlo} are shown (b) for pQCD predictions
  corrected to the particle level, (c) for three parton-shower event
  generator models, and (d) for two event generators matching
  matrix-elements to a parton shower. The scale uncertainties were
  evaluated by varying the factorization and renormalization scales by
  a factor of two.}
\end{center}
\end{figure*}

\begin{figure*}
\begin{center}
  \includegraphics*[width=\linewidth]{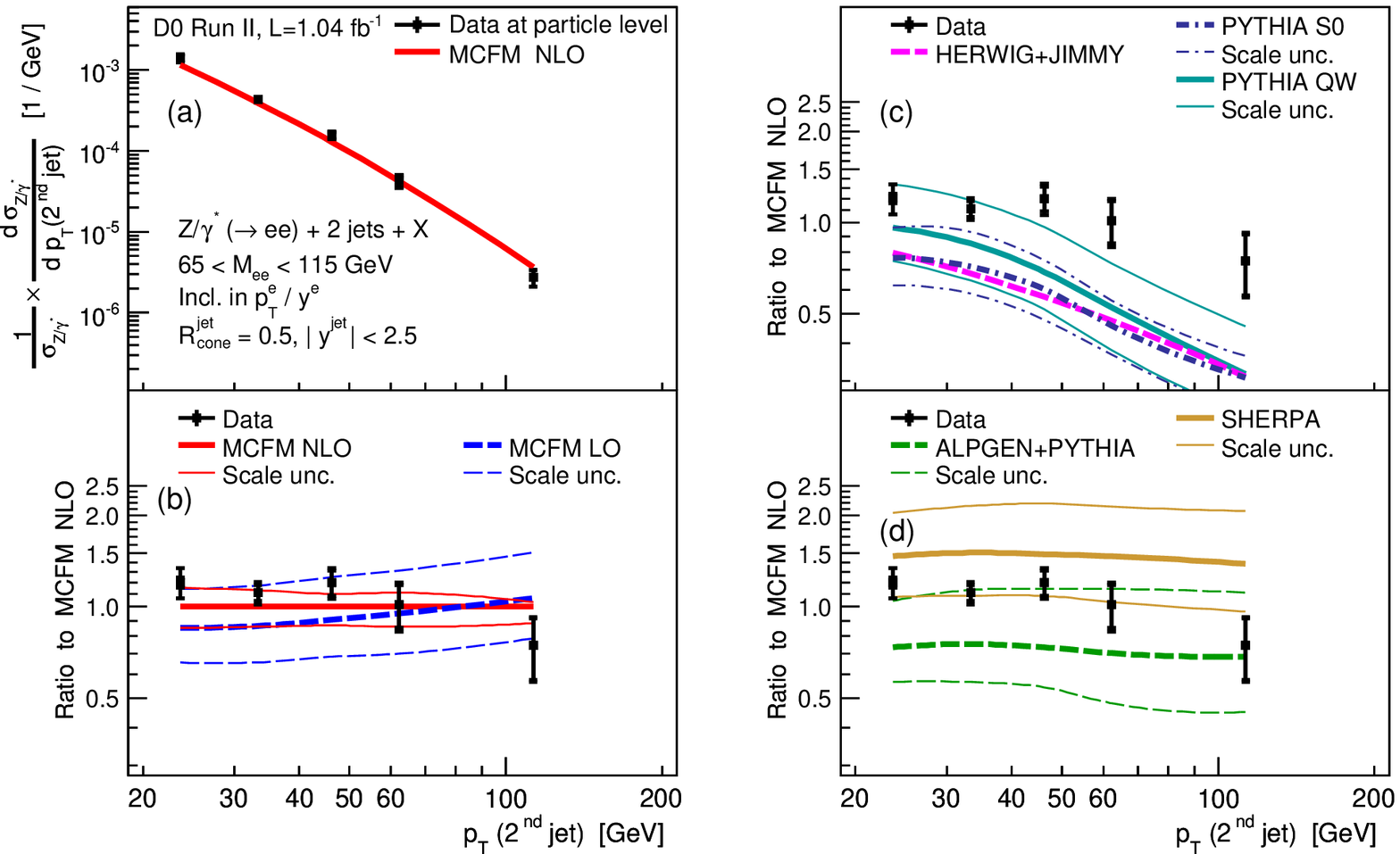}
  \caption{\label{fig2} (a) The measured distribution of
  $\frac{1}{\sigma_{Z/\gamma^*}}\times \frac{d\sigma}{d \pt({\rm
  jet})}$ for the second jet in $Z/\gamma^*+2\;$jets+$X$ events,
  compared to the predictions of {\sc mcfm nlo}. The ratios of data
  and theory predictions to {\sc mcfm nlo} are shown (b) for pQCD
  predictions corrected to the particle level, (c) for three
  parton-shower event generator models, and (d) for two event
  generators matching matrix-elements to a parton shower. The scale
  uncertainties were evaluated by varying the factorization and
  renormalization scales by a factor of two.}
\end{center}
\end{figure*}

\begin{figure*}
\begin{center}
  \includegraphics*[width=\linewidth]{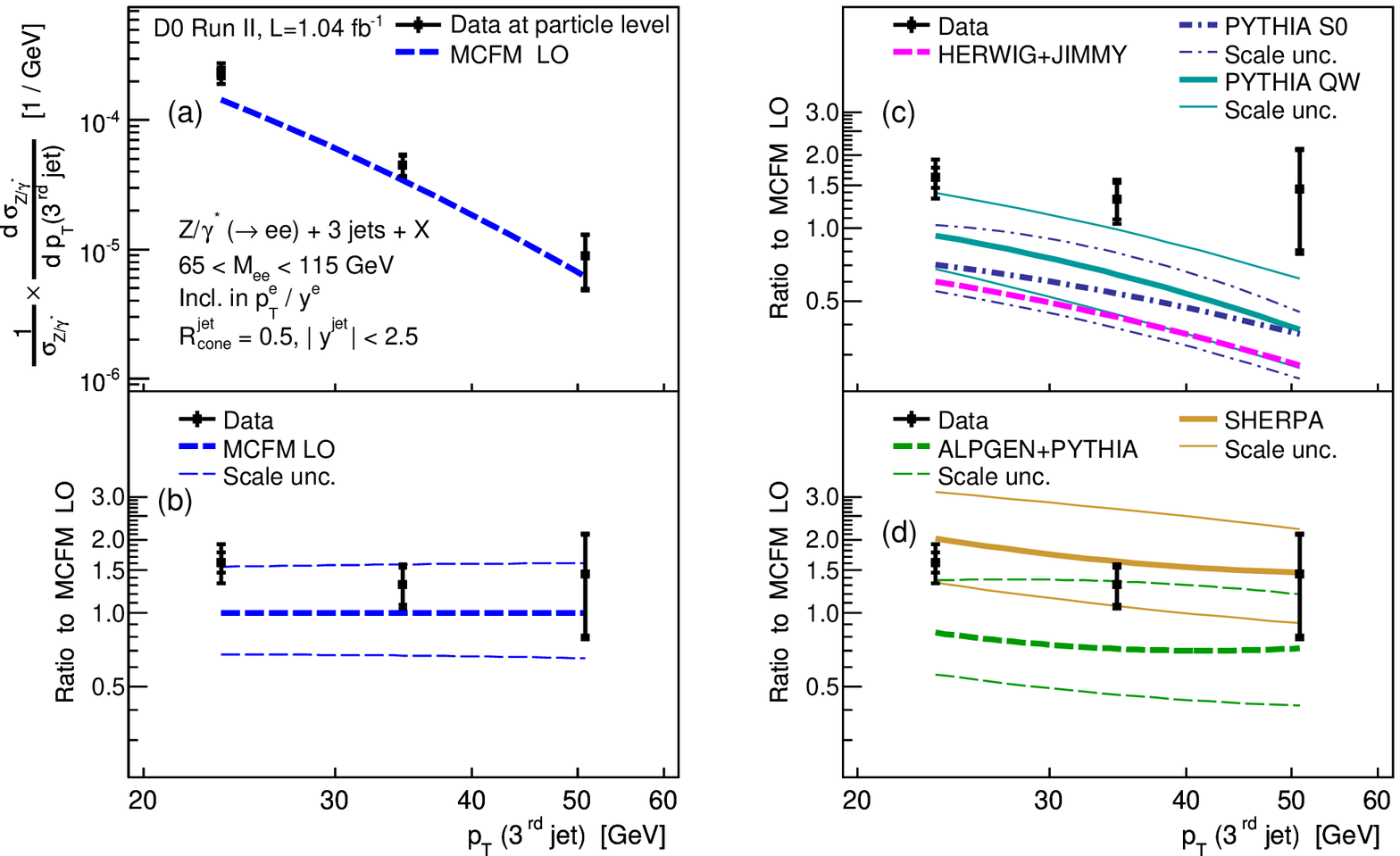}
  \caption{\label{fig3} (a) The measured distribution of
  $\frac{1}{\sigma_{Z/\gamma^*}}\times \frac{d\sigma}{d \pt({\rm
  jet})}$ for the third jet in $Z/\gamma^*+3\;$jets+$X$ events,
  compared to the predictions of {\sc mcfm lo}. The ratios of data and
  theory predictions to {\sc mcfm nlo} are shown (b) for pQCD
  predictions corrected to the particle level, (c) for three
  parton-shower event generator models, and (d) for two event
  generators matching matrix-elements to a parton shower. The scale
  uncertainties were evaluated by varying the factorization and
  renormalization scales by a factor of two.}
\end{center}
\end{figure*}

The measurements presented above are compared with the predictions of
several different theoretical models. For each model, the predicted
jet \pt~spectra are normalized to the predicted inclusive
$Z/\gamma^*(\rightarrow e^+e^-)+X$ cross section. All predictions were
generated using the CTEQ 6.1M~\cite{cteq61m} parton density functions
(PDFs) and the two-loop formula for the evolution of the strong
coupling constant ($\alpha_S$). For the first and second jets the NLO
{\sc mcfm} predictions have been taken as the reference prediction;
for the third jet, the leading-order (LO) {\sc mcfm} prediction plays
this role. The measurements and all theoretical predictions are
presented as ratios with respect to the reference prediction.

The fully corrected measurements are summarized in
Tables~\ref{tab:results_jpt0}--\ref{tab:results_jpt2} and graphically
represented in Figs.~\ref{fig1}--\ref{fig3}. The data points are shown
with statistical uncertainties (inner uncertainty bars) as well as
with statistical and systematic uncertainties combined in quadrature
(outer bars). Each data point is placed at the \pt~value where the
theoretical differential cross section is equal to the average cross
section within the bin~\cite{wyatt}.

The first comparison was performed between data and the pQCD
predictions from {\sc mcfm} v5.3 at NLO for the two leading jets, and
at LO for the three leading jets. The central predictions were defined
using factorization and renormalization scales $\mu_F = \mu_R =
\sqrt{M_Z^2 + \ptzTwo}$, with $M_Z$ and $\ptz$ denoting the mass and
transverse momentum of the $Z/\gamma^*$ boson. The sensitivity of the
predicted cross sections to the choice of $\mu_F$ and $\mu_R$ was
tested by varying their values up and down from the nominal value by a
factor of two.  The {\sc mcfm} predictions were multiplied by
correction factors accounting for multiple parton interactions
($C_{\rm MPI}$) and hadronization ($C_{\rm Had}$) before being
compared to the measurements.

The correction factors $C_{\rm MPI}$ and $C_{\rm Had}$ were estimated
using inclusive $Z/\gamma^*$($\rightarrow e^+e^-$) event samples
generated with {\sc pythia} v6.416 using Tune QW~\cite{tuneQW}, {\sc
pythia} v6.416 using Tune S0~\cite{tuneS0}, {\sc herwig} v6.510 + {\sc
jimmy} v4.31~\cite{jimmy}, {\sc alpgen} v2.13 + {\sc pythia} v6.325
using Tune QW, and {\sc sherpa} v1.1.1. The central values quoted in
Tables~\ref{tab:partonToHadron1}--\ref{tab:partonToHadron3} correspond
to the predictions of {\sc pythia} Tune QW. The maximal upwards and
downwards differences between {\sc pythia} Tune QW and the other four
models are quoted as systematic uncertainties.

\begin{table}
\caption{Correction factors for multiple parton interactions
 ($C_{\rm MPI}$) and hadronization ($C_{\rm Had}$) for
$\frac{1}{\sigma_{Z/\gamma^*}}\times \frac{d\sigma}{d \pt(1^{\rm
st}{\rm
\;jet})}$.}
\label{tab:partonToHadron1}
\begin{ruledtabular}
\begin{tabular}{r@{\hspace{-0.0mm} --\hspace{-4mm}}ccc}
\multicolumn{2}{c}{\pt~bin} & $C_{\rm MPI} \pm {\rm (stat)} \pm {\rm (sys)}$ & $C_{\rm Had} \pm {\rm (stat)} \pm {\rm (sys)}$ \\
\multicolumn{2}{c}{$[{\rm GeV}]$} & & \\
\hline
\hspace{-1.0mm}  20&28  & $1.08\pm 0.00^{+0.07}_{-0.04} $ & $0.89\pm 0.00^{+0.04}_{-0.03} $\\
\hspace{-1.0mm}  28&40  & $1.04\pm 0.00^{+0.02}_{-0.02} $ & $0.90\pm 0.00^{+0.03}_{-0.01} $\\
\hspace{-1.0mm}  40&54  & $1.02\pm 0.00^{+0.01}_{-0.01} $ & $0.90\pm 0.00^{+0.02}_{-0.00} $\\
\hspace{-1.0mm}  54&73  & $1.02\pm 0.01^{+0.00}_{-0.02} $ & $0.92\pm 0.01^{+0.01}_{-0.03} $\\
\hspace{-1.0mm}  73&95  & $1.01\pm 0.01^{+0.03}_{-0.01} $ & $0.93\pm 0.01^{+0.01}_{-0.02} $\\
\hspace{-1.0mm}  95&120 & $1.02\pm 0.02^{+0.00}_{-0.03} $ & $0.91\pm 0.02^{+0.03}_{-0.00} $\\
\hspace{-1.0mm} 120&154 & $1.04\pm 0.03^{+0.00}_{-0.07} $ & $0.92\pm 0.02^{+0.05}_{-0.03} $\\
\hspace{-1.0mm} 154&200 & $1.03\pm 0.05^{+0.02}_{-0.06} $ & $0.91\pm 0.04^{+0.04}_{-0.06} $\\
\hspace{-1.0mm} 200&300 & $1.01\pm 0.09^{+0.04}_{-0.05} $ & $0.92\pm 0.08^{+0.05}_{-0.06} $\\

\end{tabular}
\end{ruledtabular}
\end{table}

\begin{table}
\caption{Correction factors for multiple parton interactions
 ($C_{\rm MPI}$) and hadronization ($C_{\rm Had}$) for
$\frac{1}{\sigma_{Z/\gamma^*}}\times \frac{d\sigma}{d \pt(2^{\rm nd}{\rm
\;jet})}$.}
\label{tab:partonToHadron2}
\begin{ruledtabular}
\begin{tabular}{r@{\hspace{-0.0mm} --\hspace{-5.5mm}}ccc}
\multicolumn{2}{c}{\pt~bin} & $C_{\rm MPI} \pm {\rm (stat)} \pm {\rm (sys)}$ & $C_{\rm Had} \pm {\rm (stat)} \pm {\rm (sys)}$ \\
\multicolumn{2}{c}{$[{\rm GeV}]$} & & \\
\hline

\hspace{0.0mm} 20&28  & $1.15\pm 0.01^{+0.06}_{-0.10} $ & $0.81\pm 0.01^{+0.07}_{-0.00} $\\
\hspace{0.0mm} 28&40  & $1.10\pm 0.01^{+0.00}_{-0.07} $ & $0.83\pm 0.01^{+0.05}_{-0.00} $\\
\hspace{0.0mm} 40&54  & $1.07\pm 0.02^{+0.00}_{-0.06} $ & $0.85\pm 0.01^{+0.06}_{-0.00} $\\
\hspace{0.0mm} 54&73  & $1.04\pm 0.03^{+0.00}_{-0.07} $ & $0.87\pm 0.03^{+0.07}_{-0.01} $\\
\hspace{0.0mm} 73&200 & $1.05\pm 0.05^{+0.00}_{-0.08} $ & $0.83\pm 0.04^{+0.18}_{-0.00} $\\

\end{tabular}
\end{ruledtabular}
\end{table}

\begin{table}
\caption{Correction factors for multiple parton interactions
 ($C_{\rm MPI}$) and hadronization ($C_{\rm Had}$) for
$\frac{1}{\sigma_{Z/\gamma^*}}\times \frac{d\sigma}{d \pt(3^{\rm rd}{\rm
\;jet})}$.}
\label{tab:partonToHadron3}
\begin{ruledtabular}
\begin{tabular}{r@{\hspace{-0.0mm} --\hspace{-5.5mm}}ccc}
\multicolumn{2}{c}{\pt~bin} & $C_{\rm MPI} \pm {\rm (stat)} \pm {\rm (sys)}$ & $C_{\rm Had} \pm {\rm (stat)} \pm {\rm (sys)}$ \\
\multicolumn{2}{c}{$[{\rm GeV}]$} & & \\
\hline

\hspace{0.0mm} 20&28 & $1.15\pm 0.02^{+0.00}_{-0.07} $ & $0.76\pm 0.01^{+0.08}_{-0.00} $\\
\hspace{0.0mm} 28&44 & $1.10\pm 0.03^{+0.05}_{-0.04} $ & $0.81\pm 0.03^{+0.05}_{-0.00} $\\
\hspace{0.0mm} 44&60 & $1.11\pm 0.10^{+0.04}_{-0.10} $ & $0.74\pm 0.07^{+0.19}_{-0.00} $\\

\end{tabular}
\end{ruledtabular}
\end{table}

Both the NLO and the LO {\sc mcfm} predictions are in agreement with
the measurements within the experimental and theoretical uncertainties
(Figs.~\ref{fig1}--\ref{fig3}). At NLO, varying the scales up (down)
by a factor of two changes the normalized \ptjet~spectrum down (up) by
factor of $\approx$1.1 for the leading jet, compared to a factor
$\approx$1.2 for LO. For the second jet, the factors are $\approx$1.1
(NLO) and $\approx$1.4 (LO), and for the third $\approx$1.6
(LO). These numbers illustrate the improved predictive power of the
NLO computation as compared with the LO one. The uncertainties of the
{\sc mcfm} predictions due to the PDFs were evaluated using the
Hessian method. For the two leading jets, they vary from $5\%$ at low
\pt~to $10\%$ at high \pt, and for the third jet they are found to be
(5--15)\%.

Next, we compare the predictions of {\sc pythia} v6.416 and {\sc
  herwig} v6.510 + {\sc jimmy} v4.31 with the measurements. These
event generators describe jets through a parton shower using the
approximation that parton emissions are soft or collinear. For the
hard $\qqbar \rightarrow Z/\gamma^*$ scattering, both generators use
$\mu_F = M_Z$. For the parton shower, theoretical arguments favor the
choice $\mu_R = a \times p_T^{\rm rel}$, with $p_T^{\rm rel}$ being
the relative transverse momentum between the daughter partons in each
$1\rightarrow 2$ parton splitting~\cite{KtShowerScale}. This choice of
$\mu_R$ is adopted in both {\sc herwig} and {\sc pythia}, with $a=1.0$
being used in {\sc herwig} and in the final-state shower of {\sc
  pythia}. For the initial-state shower, {\sc pythia} using Tune QW
(Tune S0) sets $a=\sqrt{0.2}$ ($1.0$). Both {\sc herwig} and {\sc
  pythia} reweight the leading parton shower emission to reproduce
$Z/\gamma^*$+jet LO matrix-element computations~\cite{matching}. For
the leading jet, {\sc pythia} using Tune QW shows a more steeply
falling spectrum than observed in data (Fig.~\ref{fig1}). The
prediction of {\sc herwig} + {\sc jimmy} shows good agreement with
data at low \ptjet, but resembles {\sc pythia} Tune QW at high
\ptjet. The change of slope around $\ptjet=50$~GeV can be traced back
to the matrix-element correction algorithm in {\sc
  herwig}~\cite{seymour}. Comparisons to the measurements of the
sub-leading jets (Figs.~\ref{fig2}--\ref{fig3}) show that {\sc pythia}
using Tune QW and {\sc herwig} predict more steeply falling
\ptjet~spectra than observed in data, in agreement with expectations
based on the limited validity of the soft/collinear approximation of
the parton shower. A newer {\sc pythia} model with a \pt-ordered
parton shower, using Tune S0, gives a good description of the leading
jet, but shows no improvement for the second or third jet. For the two
{\sc pythia} models, samples were generated with $\mu_F$ and $\mu_R$
being varied up and down from the nominal value by a factor of two. As
expected, decreasing $\mu_F$ and $\mu_R$ increases the predicted
amount of events with one or more jets. The slopes of the predicted
distributions do not change significantly as the scales are varied.

Finally, we show comparisons with the {\sc alpgen} v2.13 + {\sc
pythia} v6.325 and {\sc sherpa} v1.1.1 event generators. Both
generators combine tree-level matrix elements with parton showers
\cite{ckkw,mlm1,mlm2}, thereby utilizing matrix elements also for
sub-leading jets.  For the central {\sc alpgen+pythia} prediction, the
factorization scale is given by $\mu_F = \sqrt{M_Z^2 +
\ptzTwo}$, whereas the renormalization scale is defined individually 
for each parton splitting using the CKKW prescription~\cite{ckkw}. For
{\sc sherpa}, both the factorization and the renormalization scales
are given by the CKKW prescription. For all three \ptjet~spectra, {\sc
alpgen+pythia} predicts lower production rates than observed in data,
but the shapes of the spectra are well described. {\sc sherpa}
predicts a less steeply falling leading \ptjet~spectrum than seen in
data, leading to disagreements above $40$~GeV. For the sub-leading
\ptjet~spectra, {\sc sherpa} predicts higher production rates than
observed in data, but the shapes are well described. In agreement with
Ref.~\cite{mlm2}, both {\sc alpgen+pythia} and {\sc sherpa} are found
to show a sensitivity to the choice of scales which is similar to that
of a LO pQCD prediction, reflecting a limited predictive power. For
the leading jet at $\pt=100$~GeV, the prediction of {\sc sherpa} with
both scales shifted down by a factor of two is about three times
higher than the {\sc alpgen+pythia} prediction with both scales
shifted up. This reflects both the size of the scale uncertainties and
the difference in the central prediction between the two event
generators. For {\sc alpgen+pythia}, good and simultaneous agreement
with data for all three leading jets is achieved through scaling
$\mu_F$ and $\mu_R$ down by a factor of two from the default
values. For {\sc sherpa}, an improved description of data is achieved
by scaling $\mu_F$ and $\mu_R$ up by factor of two, but remaining
disagreements with the measurements are seen for the leading jet below
$\sim 40$~GeV.

In summary, we have presented new measurements of differential cross
sections for $Z/\gamma^*(\rightarrow e^+e^-)+$jets$+X$ production in
\ppbar~collisions at a center-of-mass energy of $1.96$ TeV
using a data sample recorded by the \dzero~detector corresponding to
$1.04\pm 0.06~\mathrm{fb}^{-1}$. The measurements are binned in the
\pt~of the $N^{\rm th}$ jet, using events containing at least $N=1$,
$2$,~or $3$~jets, and are normalized to the measured inclusive
$Z/\gamma^*(\rightarrow e^+e^-)+X$ cross section. Predictions of {\sc
mcfm} at NLO, corrected to the particle level, are found to be in good
agreement with data and have a significantly smaller scale uncertainty
than {\sc mcfm} at LO. The parton-shower based {\sc herwig} and {\sc
pythia} Tune QW event generator models show significant disagreements
with data which increase with \ptjet~and the number of jets in
events. The newer \pt-ordered shower model in {\sc pythia} gives a
good description of the leading jet, but shows no improvement over the
old model for the sub-leading jets. The {\sc sherpa} and {\sc
alpgen+pythia} generators show an improved description of data as
compared with the parton-shower-based generators. {\sc alpgen+pythia}
gives a good description of the shapes of the \ptjet~spectra, while
predicting lower production rates than observed in data. {\sc sherpa}
predicts higher production rates and a less steeply falling
\ptjet~spectrum for the leading jet than observed in data. For {\sc
alpgen+pythia}, the factorization and renormalization scales can be
chosen so that a good, simultaneous agreement with data is achieved
for all three leading jets. For {\sc sherpa}, a similar level of
agreement is achieved for the sub-leading jets, but some disagreements
remain for the shape of the leading \ptjet~spectrum. Since the
presented measurements are fully corrected for instrumental effects,
they can be used for testing and tuning of present and future event
generator models.

\input acknowledgement_paragraph_r2.tex
\end{document}

%% file: list_of_authors_r2.tex
%
\author{V.M.~Abazov$^{36}$}
\author{B.~Abbott$^{74}$}
\author{M.~Abolins$^{64}$}
\author{B.S.~Acharya$^{29}$}
\author{M.~Adams$^{50}$}
\author{T.~Adams$^{48}$}
\author{E.~Aguilo$^{6}$}
\author{M.~Ahsan$^{58}$}
\author{G.D.~Alexeev$^{36}$}
\author{G.~Alkhazov$^{40}$}
% ...

\affiliation{\vspace{0.1 in}(The D\O\ Collaboration)\vspace{0.1 in}}
% ...
\affiliation{$^{6}$University of Alberta, Edmonton, Alberta, Canada;
                Simon Fraser University, Burnaby, British Columbia, Canada;
                York University, Toronto, Ontario, Canada and
                McGill University, Montreal, Quebec, Canada}
% ...
\affiliation{$^{29}$Tata Institute of Fundamental Research, Mumbai, India}
% ...
\affiliation{$^{36}$Joint Institute for Nuclear Research, Dubna, Russia}
% ...
\affiliation{$^{40}$Petersburg Nuclear Physics Institute,
                St. Petersburg, Russia}
% ...
\affiliation{$^{48}$Florida State University, Tallahassee, Florida 32306, USA}
% ...
\affiliation{$^{50}$University of Illinois at Chicago,
                Chicago, Illinois 60607, USA}
% ...
\affiliation{$^{58}$Kansas State University, Manhattan, Kansas 66506, USA}
% ...
\affiliation{$^{64}$Michigan State University,
                East Lansing, Michigan 48824, USA}
% ...
\affiliation{$^{74}$University of Oklahoma, Norman, Oklahoma 73019, USA}
% ...

%% file: acknowledgement_paragraph_r2.tex
%
We thank the staffs at Fermilab and collaborating institutions, 
and acknowledge support from the 
DOE and NSF (USA);
CEA and CNRS/IN2P3 (France);
FASI, Rosatom and RFBR (Russia);
CNPq, FAPERJ, FAPESP and FUNDUNESP (Brazil);
DAE and DST (India);
Colciencias (Colombia);
CONACyT (Mexico);
KRF and KOSEF (Korea);
CONICET and UBACyT (Argentina);
FOM (The Netherlands);
STFC (United Kingdom);
MSMT and GACR (Czech Republic);
CRC Program, CFI, NSERC and WestGrid Project (Canada);
BMBF and DFG (Germany);
SFI (Ireland);
The Swedish Research Council (Sweden);
CAS and CNSF (China);
and the
Alexander von Humboldt Foundation (Germany).

%% file: zjets_plb_resubmission.bbl
\begin{thebibliography}{99}
\bibitem{pseudorapidity} Pseudorapidity is defined as
$\eta=-\ln[\tan(\theta/2)]$, with $\theta$ being the polar angle
measured relative to the proton beam direction.
\bibitem{rapidity} Rapidity is defined as $y = \ln [(E+p_z)/(E-p_z)]$,
where $E$ is the energy and $p_z$ is the momentum along the proton
beam direction.
\end{thebibliography}
